\documentstyle[pra, aps]{revtex}
\begin{document}
\title{Aspects of mutually unbiased bases in odd prime power dimensions}
\author{S. Chaturvedi\thanks{email: scsp@uohyd.ernet.in}}
\address{ School of Physics, University of Hyderabad, Hyderabad 500046,
India}
\date{\today}
\maketitle
\begin{abstract}
We rephrase the Wootters-Fields construction  [Ann. Phys., {\bf 191}, 363 
(1989)] of a full set of mutually unbiased bases in a complex vector space 
of dimensions $N=p^r$,  where $p$ is an odd prime,  in terms of the character 
vectors of the cyclic group $G$ of order $p$. This form 
may be useful in explicitly writing  down mutually unbiased bases for 
$N=p^r$.
\end{abstract}
\maketitle
\vskip0.5cm
In a complex vector space  of dimension ${N}$, by  
a full set of mutually unbiased bases (MUB's)  we mean  a set of $N+1$ 
orthonormal bases such that the modulus square of the scalar product of 
any member of one basis with any member of any other basis is equal to $1/N$. 
If we take $e^{(\alpha, k)}$ to denote the $k^{th}$ vector in the 
$\alpha^{th}$ orthonormal basis, then having a full set of MUB's amounts to 
having a collection $e^{(\alpha, k)}~;~\alpha=0,1,\cdots,N~;~k=0,1,\cdots,N-1$ 
of $N(N+1)$, $N$-dimensional complex vectors satisfying      
\begin{eqnarray}
{|<e^{(\alpha,k)},e^{(\alpha^\prime,k^\prime)}>|^2}
&\equiv & {|\sum_{l=0}^{N-1} (e_{l}^{(\alpha,k)})^*
(e_{l}^{(\alpha^\prime,k^\prime)})|^2} \nonumber\\
&=&{\delta^{\alpha \alpha^\prime}\delta^{kk^\prime}
+\frac{1}{N}(1-\delta^{\alpha \alpha^\prime})~;~\alpha,\alpha^\prime 
=0,1,\cdots,N~;~k,k^\prime=0,1,\cdots,N-1.}
\end{eqnarray}
Here  $e_{l}^{(\alpha,k)}$ denotes the $l^{th}$ component of the $k^{th}$ 
vector belonging to the $\alpha^{th}$ orthonormal basis.  
Mutually unbiased bases thus  generalise the properties of the eigenvectors 
of the familiar Pauli matrices $\sigma_x, \sigma_y, \sigma_z$.

Though the notion of a pair of mutually unbiased bases as corresponding to 
a pair of `maximally non commuting' measurements was  introduced by 
Schwinger \cite{schwinger} as early as 1960, explicit construction of a full 
set of MUB's for $N=p$ was first given by Ivanovic\cite{ivan} and 
later by Wootters\cite{woot1}. Subsequently Wootters and Fields\cite{woot2} 
extended the construction in \cite{woot1} to the case $N=p^r$ by making use 
of the properties of Galois fields\cite{lidl}. In this work, Wootters and 
Fields, also clearly brought out the relevance of the MUB's for an optimal 
determination of the density matrix of an ensemble. It is this aspect of 
 MUB's which underlies their usefulness in quantum estimation theory and 
in quantum cryptography \cite{peres}. A recent work by Bandyopadhyay et al 
\cite{Som} contains, among other interesting results,  an explicit 
 construction of the unitary matrices (analogues of the Pauli matrices in 
$N=p^r$ dimensions) whose eigenvectors provide a full set of MUB's. 

The purpose of this Brief Report is to show that, by exploiting certain 
freedom inherent to the Wootters-Fields construction, the task of explicitly 
writing down the the full set of MUB's in odd prime power dimensions can 
be considerably simplified.     
   
We begin by noting  that for any $N$, one of the $N+1$ orthonormal bases, say, 
the one corresponding to $\alpha=N$ may always be chosen to be the 
standard basis
\begin{equation}
e_{l}^{(N,k)} = \delta_{lk}~,~l,k = 0,1,\cdots,N-1.
\end{equation}
and we can,  therefore, confine ourselves only to the remaining $N$ 
orthonormal 
bases $e^{(m,k)}$ with both $m$ and $k$ running over $0,1,\cdots,N-1$. 
These, of course, must not only be unbiased with respect to each 
other but must also be  unbiased with respect to the standard basis. 
The latter requirement implies  that $|e_{l}^{(m,k)}|$ should be equal to 
$1/\sqrt{N}$ for all $m,k,l$.
 
As noted above, explicit construction of a full set of MUB's has so far 
been possible for $N=p^r,~ p~{\rm~ a~ prime}$. For the case when $p$ is 
odd, one has \cite{woot2}
\begin{equation}
e_{{\underline l}}^{({\underline m},{\underline k})}
=\frac{1}{\sqrt{N}}\omega^{{\rm Tr}[{\bf m}{\bf l}^2+{\bf k}{\bf l}]}~~;~~
\omega = e^{2\pi i/p}.
\label{mub}
\end{equation}
Here the symbols ${\underline m}, {\underline k}, {\underline l}$
which label bases, vectors in a given basis, and components of a given 
vector in a given basis respectively, stand for $r$-dimensional arrays 
$(m_0,m_1,\cdots,m_{r-1})$ etc whose components take values in 
the set ${0,1,2,\cdots,p-1}$ i.e. in the field ${\cal Z}_p$. 
Their boldfaced counterparts ${\bf m}, 
{\bf k},{\bf l}$ which appear on the rhs of $(\ref{mub})$ belong to the 
Galois field $GF(p^r)$ i.e. they  denote polynomials 
in $x$ of degree $r$ whose components in the basis $1,x,x^2,\cdots,x^{r-1}$ 
are $(m_0,m_1,\cdots,m_{r-1})$ etc. Thus ${\underline m} \longleftrightarrow 
{\bf m} \equiv  m_0+ m_1~x+m_2~x^2+\cdots,m_{r-1}~x^{r-1}$. The variable 
$x$ is a root of a polynomial of degree $r$ with coefficients in 
${\cal Z}_p$ and irreducible in ${\cal Z}_p$ i.e. with no roots in 
${\cal Z}_p$. The trace operation on the rhs of $(\ref{mub})$ is defined as 
follows 
\begin{equation}
{\rm Tr}[{\bf m}] =  {\bf m} + {\bf m}^2+\cdots,{\bf m}^{p^r-1},
\end{equation} 
and takes elements of $GF(p^r)$ to elements of ${\cal Z}_p$. On carrying out 
the trace operation in $(\ref{mub})$ one obtains
\begin{equation}
e_{{\underline l}}^{({\underline m},{\underline k})}
=\frac{1}{\sqrt{N}}\omega^{{\underline  m}^T{\underline  q}({\underline  l})}~
\omega^{{\underline  k}^T{\underline l}}.
\label{bmu}
\end{equation}
The components of ${\underline q}({\underline l})$ are given by 
\begin{equation}
q_i({\underline l}) = {\underline l}^T~\beta_i~{\underline l}~ {\rm mod~p},
~i=0,1,2,\cdots,
 r-1 ,
\label{umb}
\end{equation}
where the $r\times r$ matrices $\beta_i~,~i=0,1,\cdots,r-1$ are obtained 
from the multiplication table of $(1, x, x^2,\cdots, x^{r-1})$  :
\begin{equation}                          
 \left(\,
\begin{array}{c}1 \\
x \\
\vdots\\
x^{r-1} \end{array}\,\right) 
\left(\begin{array}{cccc}1& x &\cdots & x^{r-1}\end{array}\right)
= \beta_0 +\beta_1~x+\beta_2 x^2+\cdots +\beta_{r-1}
~x^{r-1}.
\end{equation}
(As shown by Wootters and Fields, $(\ref{bmu})$ works for $p=2$ as well if 
we replace $\omega$ by $i$ in the first factor on the rhs 
and suspend mod $p$ operation while calculating $q_i({\underline l})$
using $(\ref{umb})$) 

We may rewrite $(\ref{bmu})$ in terms of extended arrays $({\underline m}, 
{\underline k})$ and $({\underline q}({\underline l}), {\underline l})$ as 
\begin{equation}
e_{{\underline l}}^{({\underline m},{\underline k})}
=\frac{1}{\sqrt{N}}\omega^{({\underline  m}, {\underline k})^T
({\underline  q}({\underline  l}), {\underline l})},
\label{bum}
\end{equation}
from which it is immediately obvious that if we take ${\underline l}$ to 
label the rows
and $({\underline m},{\underline k})$ to label the columns
(arranged in a lexicographical order) of an $N \times N^2$ 
matrix $e$ then the $l^{th}$ row of this matrix is given by 
\begin{equation}
 \frac{1}{\sqrt{N}} \chi^{({\underline q}({\underline l}), 
{\underline l})}\equiv 
\frac{1}{\sqrt{N}} \chi^{({ q}_0({\underline l}))}
\otimes\chi^{({ q}_1({\underline l}))}
\cdots
\otimes \chi^{({ q}_{r-1}({\underline l}))}
\otimes \chi^{({ l}_0)}
\otimes \chi^{({ l}_1)}
\otimes \cdots 
\otimes \chi^{({l}_{r-1})}  , 
\end{equation}
where $\chi^{(l)};l=0,1,\cdots, p-1$ denote the character vectors of the 
cyclic group $G$ of order $p$. The matrix $e$ contains the full set of MUB's 
- the constituent orthonormal bases are obtained by chopping this matrix 
into strips of width $N$. Of course, to write this matrix down explicitly 
one needs to work out ${\underline q}{(\underline l)}$ for each 
${\underline l}$ using $(\ref{umb})$. We now suggest a simpler way of 
achieving the same results with much less work. First, we notice that the 
rows of $e$ can be stacked on top of each other on any order. We will 
take the first row to correspond to ${\underline l}={\underline 0}$ i.e. as 
$\chi^{({\underline 0},{\underline 0})}$. To determine the remaining 
rows we proceed as follows. Choose the irreducible polynomial $f(x)$ in 
such a way that $x$ is a primitive element of $GF^*(p^r)\equiv 
GF(p^r)\backslash \{0\}$. Its powers 
$x, x^2, \cdots, x^{p^r-1}$ then give all the information we need to 
write the matrix $e$.

As as an illustration, consider the case $p=5, r=1$. Here 
$GF^*(5)= {\cal Z}_{p}^{*} = \{1, 2,3,4\}$. it is easy to see that $3$ is a 
primitive element and that its powers modulo $5$ are
\begin{equation}
3=3, 3^2 = 4, 3^3 = 2, 3^4 = 1 ,
\end{equation}
which gives $l=3 \rightarrow q(l)= 4, l=4 \rightarrow q(l)= 1,
l=2 \rightarrow q(l)= 4,l=1 \rightarrow q(l)= 1,$ and hence 
\begin{equation}
e= \frac{1}{\sqrt{5}}\left(\,
\begin{array}{c}\chi^{(0)}\otimes\chi^{(0)} \\
\chi^{(1)}\otimes\chi^{(1)} \\
\chi^{(4)}\otimes\chi^{(2)}\\
\chi^{(4)}\otimes\chi^{(3)}\\
\chi^{(1)}\otimes\chi^{(4)}\end{array}\,\right).
\end{equation} 

As another example, consider for instance $p=3, r=2$. In this case 
$f(x)= x^2+x+2$ is a polynomial of degree 2 irreducible over ${\cal Z}_3$ 
and such that $x$ is a primitive element of the multiplicative abelian group 
$GF(3^2)\backslash \{0\}$\cite{com}. Computing the powers of $x$ modulo $f(x)$ 
we obtain 
\begin{equation}
x=0+1x, x^2=1+2x, x^3=2+2x,x^4=2+0x,x^5=0+2x,x^6=2+x,x^7=1+x,x^8=1+0x,
\end{equation}
which immediately gives the ${\underline l}\rightarrow {\underline q}
({\underline l})$ correspondences. Thus $x\equiv (0,1)\rightarrow x^2 
\equiv (1,2);x^2 \equiv (1,2)\rightarrow x^4 \equiv (2,0)$ etc and we have 
\begin{equation}
e=
\frac{1}{\sqrt{9}}\left(\,
\begin{array}{c}\chi^{(0)}\otimes\chi^{(0)}\otimes\chi^{(0)}\otimes
\chi^{(0)} \\
\chi^{(1)}\otimes\chi^{(2)}\otimes\chi^{(0)}\otimes
\chi^{(1)} \\
\chi^{(1)}\otimes\chi^{(2)}\otimes\chi^{(0)}\otimes
\chi^{(2)}\\
\chi^{(1)}\otimes\chi^{(0)}\otimes\chi^{(1)}\otimes
\chi^{(0)}\\
\chi^{(2)}\otimes\chi^{(1)}\otimes\chi^{(1)}\otimes
\chi^{(1)}\\
\chi^{(2)}\otimes\chi^{(0)}\otimes\chi^{(1)}\otimes
\chi^{(2)}\\
\chi^{(1)}\otimes\chi^{(0)}\otimes\chi^{(2)}\otimes
\chi^{(0)}\\
\chi^{(2)}\otimes\chi^{(0)}\otimes\chi^{(2)}\otimes
\chi^{(1)}\\
\chi^{(2)}\otimes\chi^{(1)}\otimes\chi^{(2)}\otimes
\chi^{(2)}
\end{array}\,\right). 
\end{equation}   
Some interesting features which  become explicit  by examining the 
two factors on the rhs of $(\ref{bmu})$ are listed below:

\begin{itemize}
\item The diagonal matrices  $\Omega^{({\underline m})}$ with 
diagonal elements 
$\omega^{{\underline m}^T{\underline q}({\underline l})}$  
(${\underline l}$ taken as a row label) provide an 
$N=p^r$ dimensional unitary reducible representation of the direct product 
group $G^r=G\times G \times \cdots \times G$. This representation 
contains the trivial representation once together with half of the 
nontrivial irreducible representations which occur with multiplicity two. 

\item The diagonal matrices  ${\cal R}^{({\underline k})}$ with 
diagonal elements 
$\omega^{{\underline k}^T{\underline l}}$ 
(${\underline l}$ taken as a row label)  provide a 
$N=p^r$ dimensional unitary reducible representation of the direct product 
group $G^r=G\times G \times \cdots \times G$ which contains all the 
irreducible representations of  once (the regular representation).

\item The diagonal matrices 
$\Omega^{({\underline m})}{\cal R}^{({\underline k})}$
provide a 
$N=p^r$ dimensional unitary reducible representation of the direct product 
group $G^r\times G^r$ in which certain prescribed irreducible representations 
occur only once. This representation essentially yields the MUB's in odd prime power dimensions. 
\end{itemize}
 
To conclude, we have shown that the freedom in the choice of the 
irreducible polynomial $f(x)$ in carrying out the computations in $(6)$ 
and $(7)$ can be profitably exploited to simplify the task by 
choosing to work with an $f(x)$ whose roots are primitive elements 
of $GF^*(p^r)$.


\begin{references}
\bibitem{com} Lists of irreducible polynomials for low values of $p$ and $r$ 
together with the order of their roots may be found in\cite{lidl}.
\end{references}
\end{document}